\documentclass[journal=jacsat,manuscript=article]{achemso}
\usepackage{achemso}
\setkeys{acs}{usetitle = true}
\setkeys{acs}{maxauthors = 10}
\setkeys{acs}{etalmode=truncate}

\usepackage[version=3]{mhchem} 
\usepackage{cleveref}
\usepackage{amsmath}
\usepackage{amsbsy}
\usepackage{amssymb}

\SectionNumbersOn

\author{Marta Pelc}
\affiliation{Institute of Physics, Nicolaus Copernicus University, 87-100 Toru\'n, Poland}
\alsoaffiliation{Instituto de Ciencia de Materiales de Madrid, Consejo Superior de
Investigaciones Cient{\'{\i}}ficas, Cantoblanco, 28049 Madrid, Spain}
\email{martap@fizyka.umk.pl}

\author{Eric Su\'arez Morell}
\affiliation{Departamento de F\'{i}sica, Universidad T\'{e}cnica
Federico Santa Mar\'{i}a, Casilla 110-V, Valpara\'{i}so, Chile}
\author{Luis Brey}
\affiliation
{Instituto de Ciencia de Materiales de Madrid, Consejo Superior de
Investigaciones Cient{\'{\i}}ficas, Cantoblanco, 28049 Madrid, Spain}
\author{Leonor Chico}
\affiliation
{Instituto de Ciencia de Materiales de Madrid, Consejo Superior de
Investigaciones Cient{\'{\i}}ficas, Cantoblanco, 28049 Madrid, Spain}

\title
  {Electronic Conductance of Twisted Bilayer Nanoribbon Flakes}

\abbreviations{IR,NMR,UV}    

\begin{document}


\begin{tocentry}





\includegraphics[width=90mm]{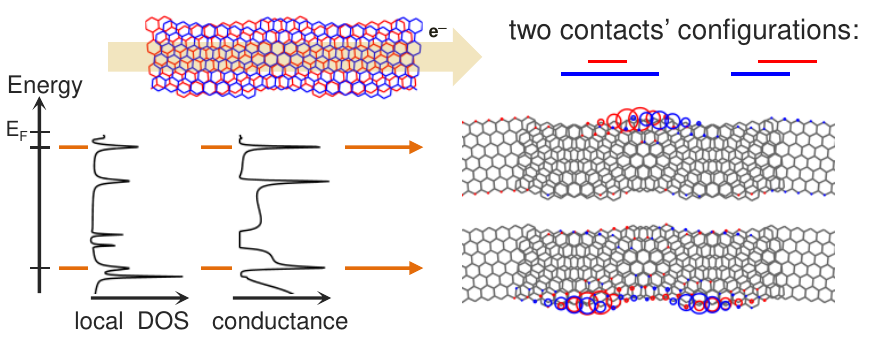}

\end{tocentry}

\begin{abstract}
We study the transport properties of a twisted bilayer graphene flake contacted by two monolayer nanoribbons which act as leads. We analyze the conductance in terms of the spectra of the bilayer nanoribbon and the monolayer contacts. The low-energy transport properties are governed by the edge states with AB stacking. 
Remarkably, the electronic conductance  in this energy region does not depend much on the relative position of the leads, in contrast with that of bilayer flakes with more symmetric stackings. We attribute this feature to the localization of these low-energy states in the AB edge regions of the flake, having a much smaller weight at the junctions between the flake and the nanoribbon leads.
\end{abstract}


\section{Introduction}
The electronic properties of graphene can be tuned by stacking a few layers of this material \cite{Guinea_2006,CastroNeto_RMP_81,Brown2012,Ling2013,pccp}, opening a way to tailor its band structure and even open a gap by the application of an electric field \cite{Latil_2006,Castro_2007,Min_2007,Mak_2009}.
Besides the number of layers, different stacking orders also result in distinct electronic and optical properties in multilayer graphene that have been experimentally probed. \cite{Cong2011,Ling2013,Wu2014}
 In fact, the low-energy band structure of bilayer graphene strongly depends on the stacking sequence. While AA stacking preserves the linear dispersion around the Fermi level, typical for a single layer \cite{Ho_2006,Liu2009,Prada_2011}, AB-stacked graphene has a parabolic dispersion relation \cite{Novoselov_2006,McCann_2006,Aoki_2007}. Furthermore, rotating the two layers is yet another way to modify the spectrum of bilayer graphene that has been also explored \cite{Lopes_2007,Campanera_2007,Latil_2007,Shallcross_2008,Varchon2008,Mele_2010,Li2010,Lopes_2012,Bistritzer,Landgraf_2013}. In this latter system, also called twisted bilayer graphene (TBG), the rotation angle between layers determines the electronic properties of the system and different electronic regimes have been identified \cite{Hicks_2011,PNAS_moire,Luican2011}. For large rotation angles, the system behaves as two noninteracting monolayers, where for small rotation angles---below 10$^{\circ}$---the Fermi velocity decreases, dropping to zero around 1$^{\circ}$\cite{Trambly_2010,SuarezMorell2010}. 
 Moreover, scanning tunneling measurements have revealed large superperiodicities in the density of states of TBG,\cite{Berger2006} described as moir\'e patterns \cite{Campanera_2007}. Regions with different types of stacking, AA and AB, as well as displaced graphene layers, called slip regions, can be identified. These areas have distinct electronic localizations depending on the energy, giving rise to inhomogeneous distributions of the electrons. In fact, for small angles a van Hove singularity in the density of states  develops at low energies \cite{Trambly_2010,Li2010}.
Moir\'e patterns have also been observed in multilayer graphene\cite{Wu2014} and in highly oriented pyrolytic graphite (HOPG) with a misoriented layer on top.\cite{SuarezMorell2015} These superperiodic structures can be additionally tuned by molecular adsorption\cite{Meng2014} and atom deposition,\cite{Zhang2014} creating new avenues for graphene structures. 

Besides the opening of a gap, the application of graphene in nanoelectronics requires the realization of nanometric-size structures for the design of devices. Graphene nanoribbons emerge as optimal candidates for the development of carbon-based components at the nanoscale, with the added possibility of modifying its electronic properties because of finite-size effects \cite{Nakada_1996,Wakabayashi_1999,Baringhaus_2014,Han_2007}. 
Indeed, the electronic properties of monolayer nanoribbons strongly depend on their width and edge shape \cite{Brey2006,Akhmerov2008,Castro_2008,Jaskolski_edges,Santos_2013,Son_2006,Han_2007}. In the case of bilayer ribbons, the role of stacking has to be considered additionally \cite{Sahu_2008,Feng_PRB_80,Lima_2009,Santos_2012}. In particular, for zigzag nanoribbons with AA stacking, the edge flat bands at the Fermi level split, while for AB-stacked zigzag nanoribbons, the edge bands are degenerate at the Fermi level. This difference can be explained in terms of the larger interlayer coupling in AA stacking, which splits the edge bands into bonding and antibonding. 

The electronic transport properties of bilayer graphene flakes have also been recently addressed for the most symmetric geometries, i.e., zigzag and armchair edges and direct (AA) and Bernal (AB) stackings \cite{Gonzalez_2011,Gonzalez_gate,Yin_2013,Orellana_2013}. 
Oscillations of the conductance with respect to the energy, flake length and interlayer coupling were reported. 
Notably, the conductance of the flakes with different lead positions (i.e., both connected to the same layer or to opposite ones) showed a complementary behavior. Such transport characteristics are relevant for the design of electromechanical devices, and deserve further exploration in more general geometries, such as moir\'e bilayer ribbons and chiral edges. Although there is no controllable way to fabricate them, twisted bilayer structures appear naturally in some growth methods \cite{Li2010,Varchon2008}. Moreover, unzipping multiwall nanotubes can be used to obtain twisted bilayer nanoribbons \cite{Kosynkin2009,Jiao2009,Jiao2011}.

In this work we study the transport properties of twisted bilayer graphene nanoribbons. By cutting a stripe from a twisted bilayer graphene, different atomic stackings may appear at the edges. Therefore, an interplay between the moir\'e pattern and the graphene edges is expected to govern the low-energy transport through these systems, as it has been recently evidenced. \cite{SuarezMorell2014}
The low-energy properties of twisted bilayer ribbons are closely related to the AB-stacked portions of the edges. This behavior is in contrast to that of bulk twisted graphene, which develops a peak of the local density of states (LDOS) at the AA-stacked regions for these energies. The increase of the local density of states is relevant for functionalization of the system, due to the enhanced reactivity of site with this character, such as folds, edges and defects \cite{Yan,Sharma2009,Denis}.

In order to investigate the electronic transport properties of twisted bilayer nanoribbons, we consider a bilayer flake connected to monolayer leads, as previously done 
for armchair and zigzag bilayer ribbons with AA and AB stackings \cite{Gonzalez_2011,Gonzalez_gate}. We analyze the appearance of localized states near the Fermi energy, and explore their origin, i.e., their relation to edge and bilayer stacking, as well as their role in the conductance of the system. 

We briefly state here our main results:\\
\noindent
(i) Different transport regimes can be identified in the system, namely, resonant transport, tunneling and antiresonances. These distinct  behaviors have been analyzed with respect to the spectrum of the flake and the nanoribbon contacts.\\
\noindent
(ii) We have identified two types of localization, one due to the moir\'e at the edges, which gives rise to mostly 
AB and AA edge states, and other due to the flake boundaries and contacts to the leads, being a finite-size effect of the whole flake. \\
\noindent
(iii) In general, the low-energy transport properties of twisted bilayer flakes are governed by AB-stacked edge states, even for wide ribbons. As the edge-localized states are confined to the corresponding regions of the edges, their weight in the lead-flake junctions is very low, so their transport properties do not depend appreciably on the 
position of the contacts for this energy range, differently to the behavior of symmetrically stacked flakes. \\
\noindent
(iv) For higher energies, finite-size, flake-localized states spread up to the leads, so the conductance is strongly dependent on the configuration of the leads and shows a  complementary behavior between the two configurations, as previously found in bilayer ribbons with symmetric (AA and AB) stackings \cite{Gonzalez_2011}.

The rest of the paper is organized as follows: Sec. \ref{sec:geometry} describes the geometry of the twisted bilayer flake and the corresponding monolayer ribbon contacts. In Sec. \ref{sec:method} we explain the tight-binding model for the bilayer and the calculation details. Sec. \ref{sec:results} presents our results and discussion. 
Finally, in Sec. \ref{sec:summary} we summarize our main results and conclusions. 

\section{\label{sec:geometry}Geometry }

In this Section we briefly describe the geometry of the system studied, a bilayer graphene ribbon with monolayer ribbon contacts. 

\subsection{\label{sec:tblnr}Twisted bilayer nanoribbon}

\begin{figure}[thpb]
      \centering
\includegraphics[width=\columnwidth]{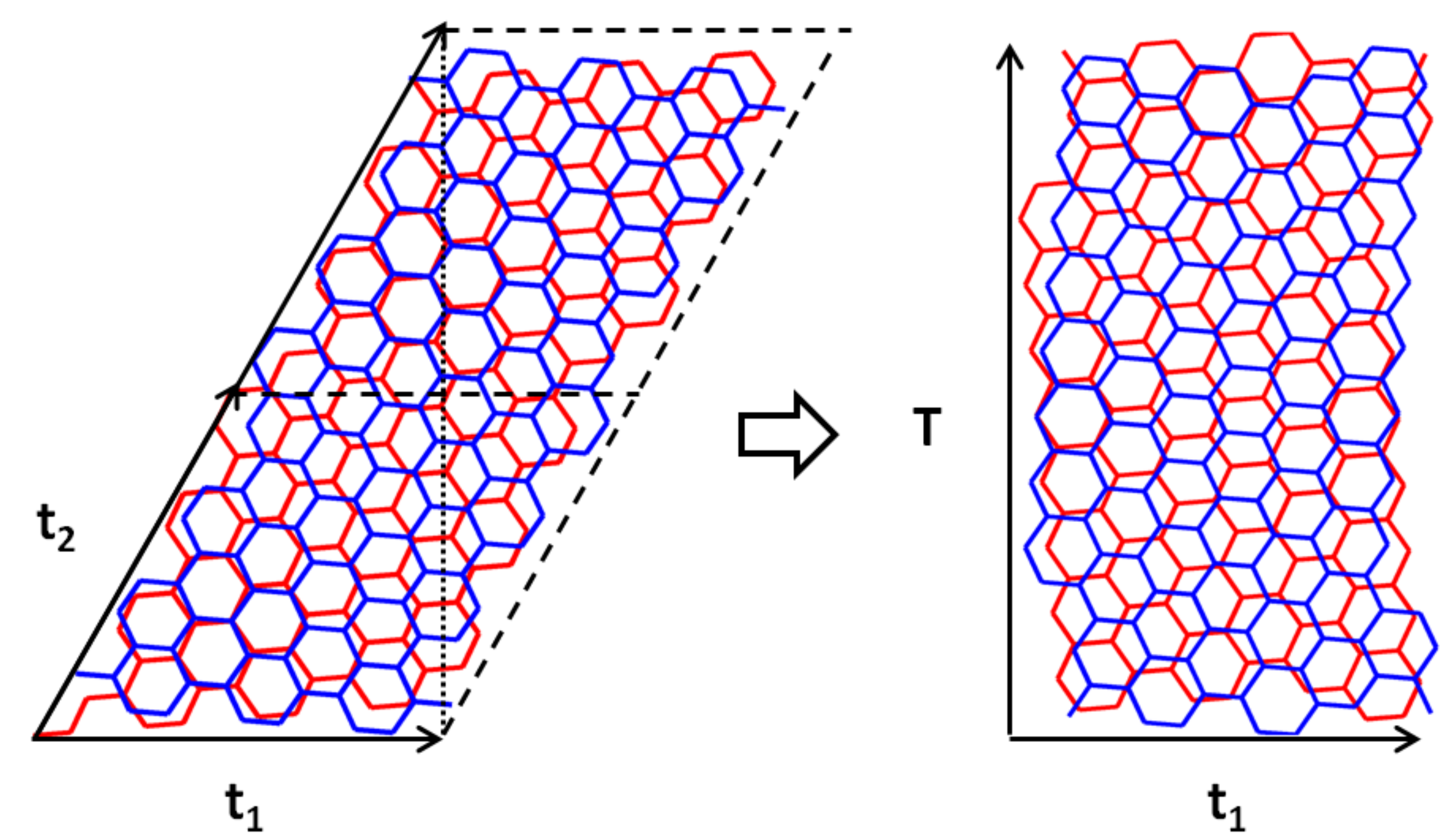}
\caption{\label{fig:uc}
(Color online). Construction of the unit cell for the $(10,1)[(4,3)]$ twisted bilayer nanoribbon.
From the unit cell spanned by $\vec{t}_{1}$ and $\vec{t}_{2}$,
we build 
a rectangular unit cell, defined by $\vec{t}_{1}$ and $\vec{T}$
with mostly zigzag egde atoms. $\vec{T}$ is the translation 
vector of the nanoribbon and its edge vector.}
   \end{figure}

 We build the unit cell (UC) of the twisted bilayer nanoribbon from that of the twisted bilayer graphene, shown in Fig. \ref{fig:uc} \cite{SuarezMorell2014}. Recall that the UC for commensurate twisted bilayer is generated by rotating one of the AB-stacked layers about an axis passing through a B atom, 
so that an atom with coordinates $\vec{r}=m\vec{a}_{1}+n\vec{a}_{2}$ is rotated to an equivalent site $\vec{t}_{1}=n\vec{a}+m\vec{a_{2}}$. Here $\vec{a}_{1}=\frac{a}{2}(\sqrt{3},-1)$ and $\vec{a}_{2}=\frac{a}{2}(\sqrt{3},1)$ are the graphene lattice vectors and $a=2.46 $ {\AA} is the graphene lattice constant. The relative rotation angle (RRA) is completely specified by the integers $n$ and $m$, namely,  $\cos\theta=(n^{2}+4nm+m^{2})/2(n^{2}+nm+m^{2})$. The unit cell of the system is defined by the vectors $\vec{t}_{1}=n\vec{a}+m\vec{a_{2}}$ and $\vec{t}_{2}=-m\vec{a}_{1}+(n+m)\vec{a}_{2}$, which span an angle
of $60^{\circ}$. Since the two integers $n$ and $m$ fully determine
the unit cell, they are usually chosen as labels for the moir\'e bilayer, $(n,m)$. We select them to fulfill the condition of $n=m+1$. 
In principle, a twisted bilayer nanoribbon can be generated by repeating the UC along the directions given by $\vec{t}_{1}$ or $\vec{t}_{2}$, but this would produce edges
with a dominant armchair component. As zigzag-terminated edges have low-energy, edge-localized states, with remarkable transport properties, we 
choose another unit cell, with a translation vector $\vec{T}=2\vec{t}_{2}-\vec{t}_{1}$ perpendicular to $\vec{t}_{1}$, so that the edge runs along $\vec{T}$. This rectangular UC is twice the size of the minimal one, 
 but with an edge vector $\vec{T}=(2m+n,n-m)$ with mostly zigzag character. For $n=m+1$ this implies that $\vec{T}=(3m+1,1)$.  We choose the edges to be minimal, i.e., with a minimum number of edge atoms, all with coordination number 2.\cite{Akhmerov2008,Jaskolski_edges} Twisted bilayer nanoribbons are labeled indicating the coordinates of the edge vector $(3m+1,1)$ and the moir\'e pattern in brackets $[(n,m)]$, that is, $(3m+1,1)[(n,m)]$.\cite{SuarezMorell2014} Although this information is redundant, it helps to identify more easily the geometry of the ribbons. We omit in this work the width index because all examples chosen herein have the same, equal to $|\vec{t}_{1}|$, so it would be always equal to 1 in units of the bilayer nanoribbon width vector $\vec{t}_1$.


\subsection{\label{sec:mod_sys} Bilayer flake with monolayer leads}

\begin{figure}[thpb]
      \centering
\includegraphics[width=\columnwidth,clip]{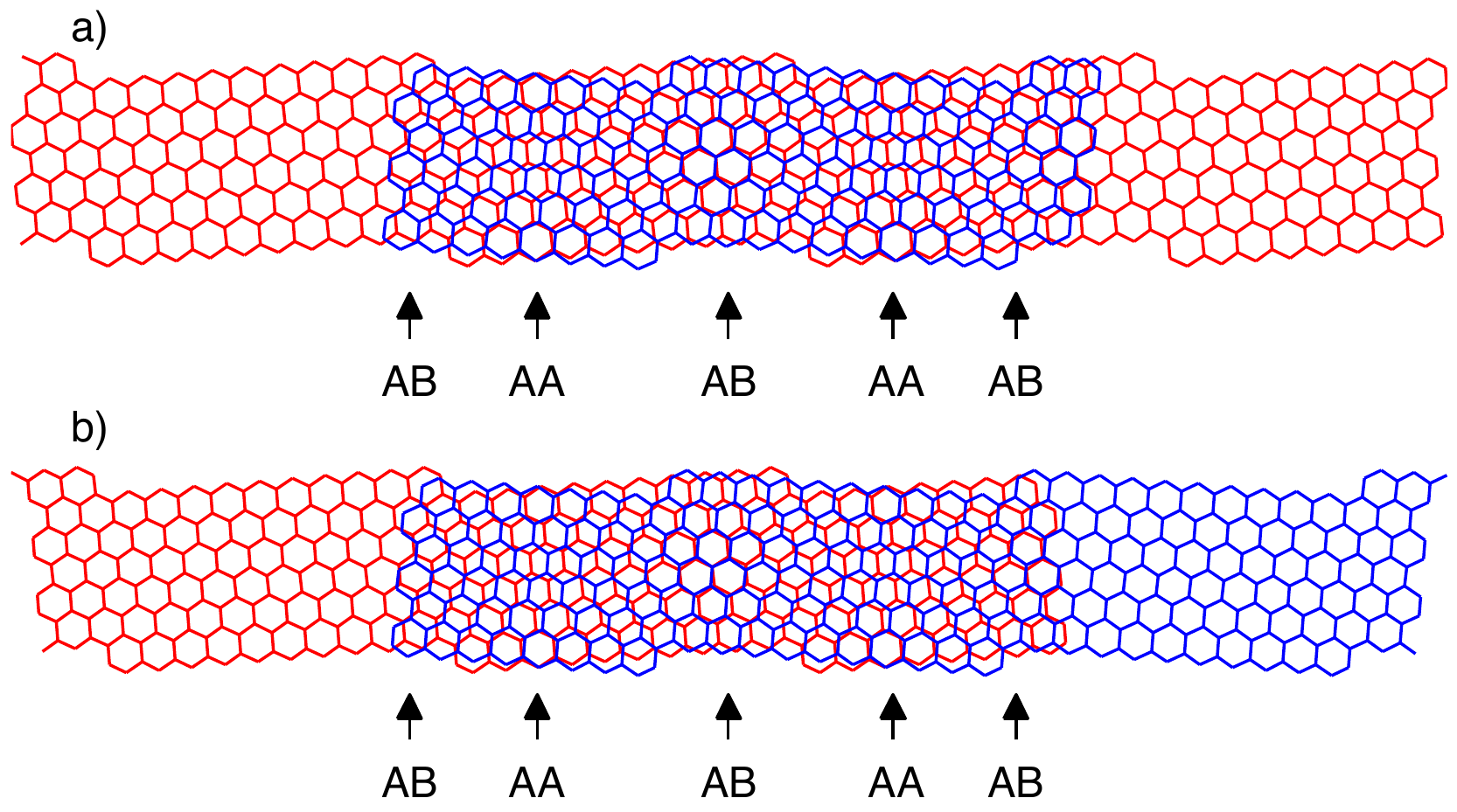}
\caption{\label{fig:qd_uc}
(Color online). Two ways of building twisted bilayer ribbon flakes (10,1)[(4,3)] with monolayer ribbon leads: (a) 1-1 or {\it island} configuration - a monolayer ribbon with a finite-size flake on top, and (b) 1-2 or {\it overlap} configuration - two overlapping monolayer ribbons which yield a twisted bilayer flake. The length of the flake is equal to two unit cells ($2|\vec{T}|)$, approximately 50 \AA, and its width $|\vec{t}_{1}|$ is about 15 \AA.}
   \end{figure}

We construct the flake by connecting the finite twisted bilayer
ribbon with semiinfinite monolayer leads with the same edge vectors and widths. 
As previously done with achiral ribbons \cite{Gonzalez_2011}, we consider two ways of devising a finite-size bilayer flake with contacts. 
First, we put a flake on top of an infinite ribbon (Fig. \ref{fig:qd_uc} (a)), so that the monolayer parts of the system play 
the role of leads. We refer to this as the 1-1  or {\it island} configuration. 

Second, the leads are connected to different layers of the flake, so the system can be viewed as two overlapping semiinfinite ribbons (Fig. \ref{fig:qd_uc} (b)). We will refer to it as the 1-2 or {\it overlap} configuration. As our interest is to elucidate the interplay of edge states and the moir\'e pattern, which produces different stackings on the edge, we avoid other sources of localization by choosing
all edges to be minimal in both configurations, without Klein atoms \cite{Klein}, coves or capes \cite{Jaskolski_edges}, and having an armchair shape at the ends of the flake. 


\section{\label{sec:method}Model and Method}

We model the system with a $p_{z}$-orbital tight-binding Hamiltonian. For the  monolayer ribbons constituting the leads we consider only the nearest-neighbor hopping parameter $\gamma_{0}=-3.16$ eV and on site energy $\epsilon_0=0$. 

For the bilayer flake,  the Hamiltonian can be written as
\begin{equation}
H=H_{1}+H_{2}+H_{12},
\end{equation}
where $H_{1}$ and $H_{2}$ are single-layer Hamiltonians
and the interlayer
Hamiltonian $H_{12}$ is given by 
\begin{equation}
H_{12}=\sum_{i,j}\gamma_{1}e^{-\beta(r_{ij}-d)}c_{i}^{\dagger}c_{j}+H.c., 
\end{equation}
where $\gamma_{1}=-0.39$ eV is the interlayer hopping, 
$r_{ij}$ is the difference between in plane coordinates of atoms $i$
and $j$ belonging to different layers, $d=3.35 $ {\AA}
is the distance between layers, and $\beta=3$ is a parameter fitted 
to  density functional theory (DFT) calculations \cite{Shallcross_2008,Latil_2007}. 
We take a cutoff for the interlayer interaction of $6a_{\rm CC}$, with  $a_{\rm CC}=1.42$ {\AA}.

We choose a tight-binding approach due to large number of atoms in the system, given that the flake alone consists of almost 600 atoms. 
Undoubtedly, calculations including dispersive forces, as in previous works in symmetric stackings \cite{Santos_2012,Lima_2009} would be of great interest, so to 
determine changes in the geometry that might lead to variations in the electronic properties, but this is 
beyond the scope of this work. 
Our choice of method and parameters has been thoroughly tested previously:
on the one hand, this tight-binding description of twisted bilayer graphene was shown to be in excellent agreement with DFT results around the Fermi energy of the system, which is the range of interest for the present work.\cite{SuarezMorell2010} On the other hand, scanning tunneling microscopy (STM) measurements of a twisted graphene layer on top of HOPG have been successfully explained with this model.\cite{SuarezMorell2015}  In general, tight-binding approaches have been extensively employed for the interpretation of experimental results in few-layer graphene. Differences in the infrared optical conductivity of gated ABC and ABA trilayer graphene\cite{Lui2011} and quantum Hall effect measurements in these systems\cite{Thiti2011} have been satisfactorily explained with tight-binding parametrizations. And with respect to twisted bilayer graphene, two of the most relevant features of this system, namely the appearance of van Hove singularities at low energies\cite{Li2010} and the renormalization of the Fermi velocity,\cite{Luican2011} have also been experimentally observed and satisfactorily modeled with tight-binding methods. 

Since the bilayer flake with leads has no
translational symmetry, we use the Green function matching method to compute the density of states and
the conductance. The
total Hamiltonian of the system is given by \cite{Datta}:
\begin{equation}
H=H_{L}+H_{R}+H_{C}+V_{LC}+V_{RC},
\end{equation}
where $H_{L}$ and $H_{R}$ are the Hamiltonians of the left (L) and right (R) lead respectively,
 single-layer ribbons in this case, $H_{C}$ is the Hamiltonian of the central
part or conductor, i.e., the twisted bilayer flake, and $V_{LC}$ and $V_{RC}$ are the connections between the conductor and the left and right leads, respectively.
The Green function of the conductor $C$ is then \cite{Datta}
\begin{equation}
\mathcal{G}_{C}(E)=(E-H_{C}-\Sigma_{L}-\Sigma_{R})^{-1}
\end{equation}
where $E$ is the energy, $\Sigma_{L}=V_{LC}g_{L}V_{LC}^{\dagger}$ and $\Sigma_{R}=V_{RC}g_{R}V_{RC}^{\dagger}$
are the leads self-energies and $g_{L,R}$ are the Green functions of the monolayer leads.
The local density of states (LDOS) is given by 
\begin{equation}
\mathrm{LDOS}(E)=-\frac{1}{\pi}{\rm Im}({\rm Tr}(\mathcal{G}_{C}(E))). 
\label{eq:ldos}
\end{equation}
In this work the trace is taken over all the nodes of the bilayer flake. In certain cases we also plot the atom-resolved LDOS (without summing over the nodes) 
in order to see the spatial distribution of the state in the conductor.

The conductance is calculated using the Landauer-B\"uttiker formalism,
\begin{equation}
G=\frac{2e^{2}}{h}T(E)=\frac{2e^{2}}{h}{\rm Tr}[\Gamma_{L}\mathcal{G}_{C}\Gamma_{R}\mathcal{G}_{C}],
\end{equation}
where $T(E)$ is the transmission function from the left to the
right lead and $\Gamma_{L,R}=i[\Sigma_{L,R}-\Sigma_{L,R}^{\dagger}]$
describe the coupling between the conductor and the $L,R$ leads.

\section{\label{sec:results}Results and discussion}

Most of our results will be shown for $(10,1)[(4,3)]$ bilayer flakes, with RRA equal to $\theta=9.43^{\circ}$ and number of atoms per unit cell $N=296$. We have also performed calculations for smaller systems, like $(4,1)[(2,1)]$, and larger systems, such as $(13,1)[(5,4)]$. However, on the one hand, for ribbons with smaller unit cells and large RRA, the moir\'e pattern is too small to observe a clear AA and AB edge stacking. On the other hand, when the flake unit cell is very large, the analysis becomes too cumbersome,  because the spectrum is much denser with numerous quasi-localized states very close in energy, although the features discussed here remain the same. The selected system, $(10,1)[(4,3)]$, is large enough to distinguish AA and AB stacking at the edges, and has a clearer spectrum, having all 
the relevant characteristics of larger ribbons.

Unless stated otherwise, the length of the flake is equal to two unit cells of the twisted bilayer ribbon. Thus we may distinguish stacking-related localization from that due to the finite size of the flake. That rises the number of atoms in a flake $(10,1)[(4,3)]$ up to $2N=592$.

\begin{figure}[thpb]
      \centering
\includegraphics[width=\columnwidth]{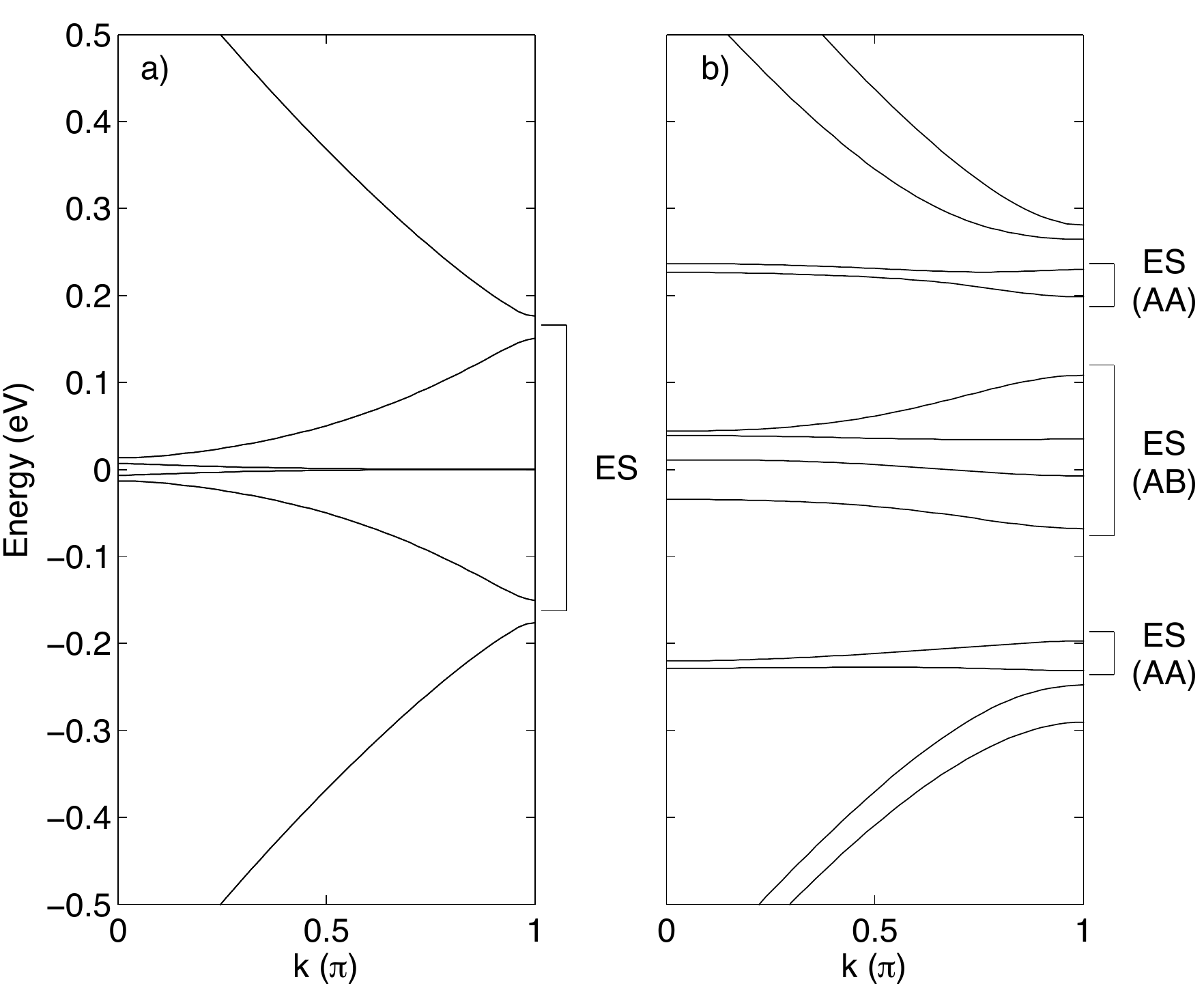}
\caption{\label{fig:media}
Band structures of the constituent media: (a) infinite monolayer ribbon (10,1) with  minimal edges; 
(b)  twisted
bilayer ribbon $(10,1)[(4,3)]$. 
 On the right sides of the band structure panels we mark with square brackets and the label ES the energy intervals for edge bands. For the twisted bilayer ribbon, AA and AB edge localized bands are correspondingly labeled. 
}
   \end{figure}

Before carrying on with the results for the flakes, it is worthy to summarize previous research on the electronic properties of the constituent media, namely, the monolayer ribbons, which play the role of contacts, and the twisted bilayer nanoribbon constituting the conductor. 
In Fig. \ref{fig:media} we plot the band structures of the corresponding {\it infinite} ribbons: (a) the monolayer chiral ribbon with edge vector $\vec{T}=(10,1)$, and (b) the  $(10,1)[(4,3)]$ twisted bilayer ribbon.
The (10,1) edge has 2 edge states \cite{Jaskolski_edges}, so the monolayer ribbon has 4 edge bands close to the Fermi energy, roughly between $-0.15$ eV and 0.15 eV.  As the ribbon is narrow, edge-localized states couple, so the edge
bands have a relatively large dispersion that decreases for wider ribbons. 


For the twisted bilayer nanoribbon we thus expect 8 edge bands, that we observe from $-0.25$ to $0.25$ eV.  In this case, there is a difference between localization depending on the stacking. States closer to the Fermi level are mostly localized on the AB-stacked regions of the edges, while edge states farther from Fermi energy, around $|E| \approx$ 0.2 eV, localize on regions with AA stacking. Such energy difference related to the type of stacking can be clearly understood considering edge states in bilayer zigzag ribbons \cite{SuarezMorell2014}. As mentioned before, edge states for AB-stacked ribbons are very close to $E_F$, whereas AA stacking splits the edge bands away from $E_F$ an amount related to the interlayer coupling. In twisted bilayer ribbons, edges have a mixture of different stackings; nonetheless, we can verify that such relation between stacking and edge bands splitting remains.

\subsection{\label{sec:island}LDOS and conductance of bilayer flakes}

\begin{figure}[thpb]
      \centering
\includegraphics[width=\columnwidth]{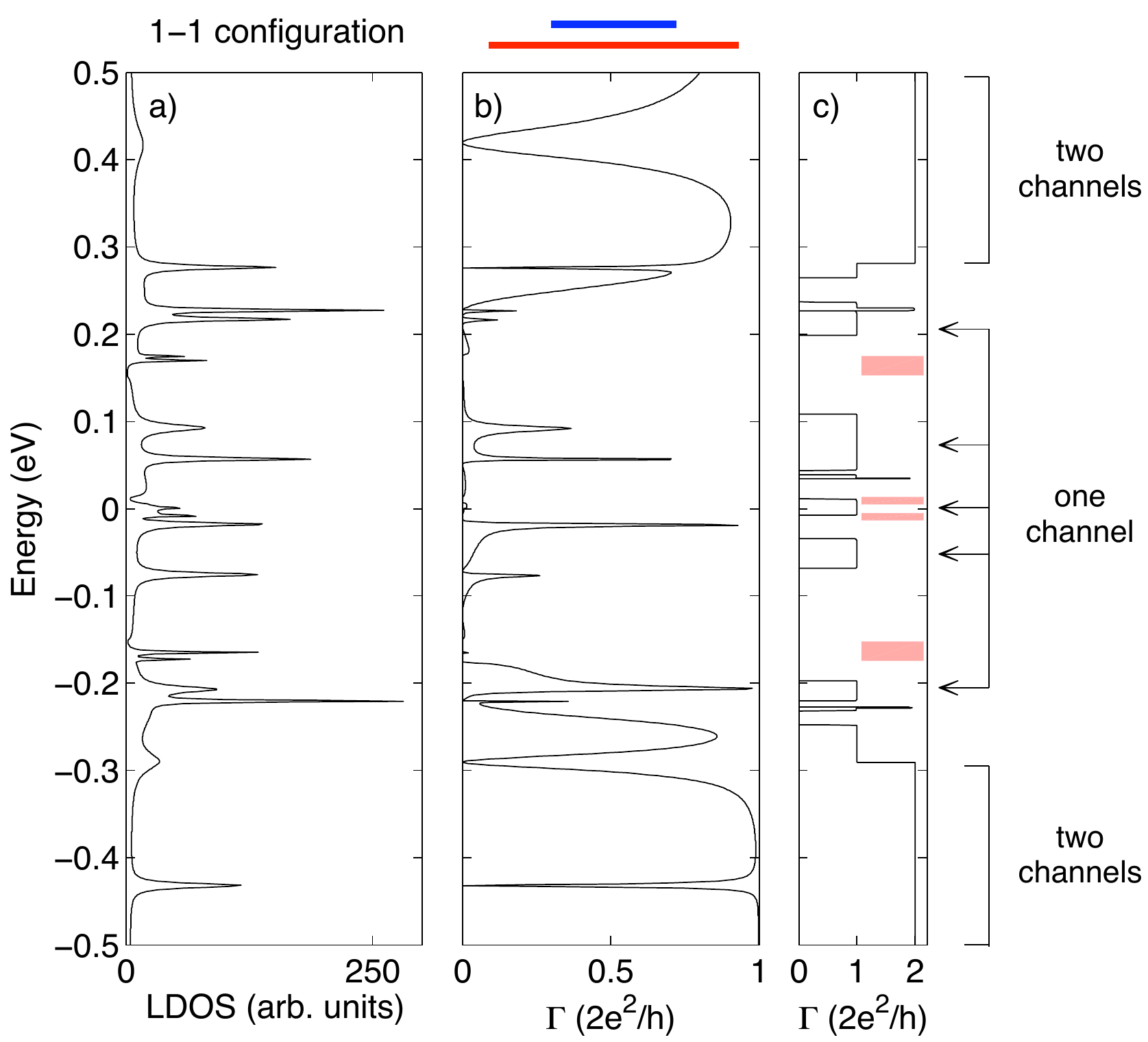}
\caption{\label{fig:is_ldos_cond} (Color online). 
(a) LDOS and (b) conductance of an island (10,1) [(4,3)] flake (configuration 1-1) with length equal to 2 unit cells. For comparison, we show (c) the conductance of the corresponding infinite twisted bilayer (10,1) [(4,3)] ribbon. The pink rectangles mark the conductance gaps of a monolayer ribbon with $\vec{T}=(10,1)$ and width equal to $\vec{t}_1$. Arrows indicate some of the main one-channel energy ranges for the bilayer ribbon. The broader two-channel energy intervals are marked with square brackets.}
   \end{figure}

First we consider the 1-1 configuration, a monolayer nanoribbon with a finite-size graphene flake on top.  
Fig. \ref{fig:is_ldos_cond} depicts the LDOS summed over all atoms in the flake along with the conductance through
this system (Eq. \ref{eq:ldos}). The rightmost panel (c) represents the conductance of the infinite bilayer flake, which allows us to have in mind the number of conductance channels in the conductor as well as the available lead states: 
pink rectangles mark the energy gaps in the infinite single layer ribbon (10,1), which otherwise has one conductance channel in the depicted energy range.  The conductance of the infinite systems is computed just by counting states. 

We can easily correlate the peaks in the LDOS 
(Fig. \ref{fig:is_ldos_cond} (a)) with distinct features in the conductance through the flake 
(Fig. \ref{fig:is_ldos_cond} (b)).  Notice that they have a remarkable energy dependence. In the region closer to $E_F$, energies $|E| < 0.15$ eV, they correspond to conductance maxima, and from the bulk band structure presented in 
Fig. \ref{fig:media}, we infer that they correspond to AB-localized edge states in the flake. 
Likewise, we expect some maxima in LDOS and conductance around $|E| \approx$ 0.2 eV to be caused by AA-localized edge states. 

In any case, these edge-localized states are coupled to the leads, yielding a non-zero and maximum conductance. 
For energies farther from $E_F$ the situation is more complex. 
If there are two available states in the conductor, antiresonances take place \cite{Orellana_2003}. As depicted in panel (c), the two-channel energy ranges happen for $|E|\gtrsim 0.3$ eV, plus three narrow spikes due to a slight band bending of edge bands. The antiresonance behavior is clearly seen at $-0.29$ eV, where the LDOS peak matches a clear conductance zero minimum. 
In these cases there is a destructive interference between the two available states in the conductor. The two LDOS peaks at $E=0.42$ eV and $E=-0.43$ eV are two clear examples of this antiresonant behavior. This behavior was also observed in graphene flakes with more symmetric stackings \cite{Gonzalez_2011}. 

The conductance of the island flake, Fig. \ref{fig:is_ldos_cond} (b), presents clear gaps that correspond to those in the monolayer leads, marked as pink bands in panel (c). Notice, however, that there is a non-negligible conductance in energy windows without available states in the bulk bilayer conductor, such as from $0.1$ to 0.2 eV or from $-0.2$ to $-0.06$ eV (see panel (c)). In these regions there are broad conductance peaks and shoulders due to tunneling through the flake. This is related to the small size of the central conductor, which is 50 \AA, but only two unit cells long.

\begin{figure}[thpb]
      \centering
\includegraphics[width=\columnwidth]{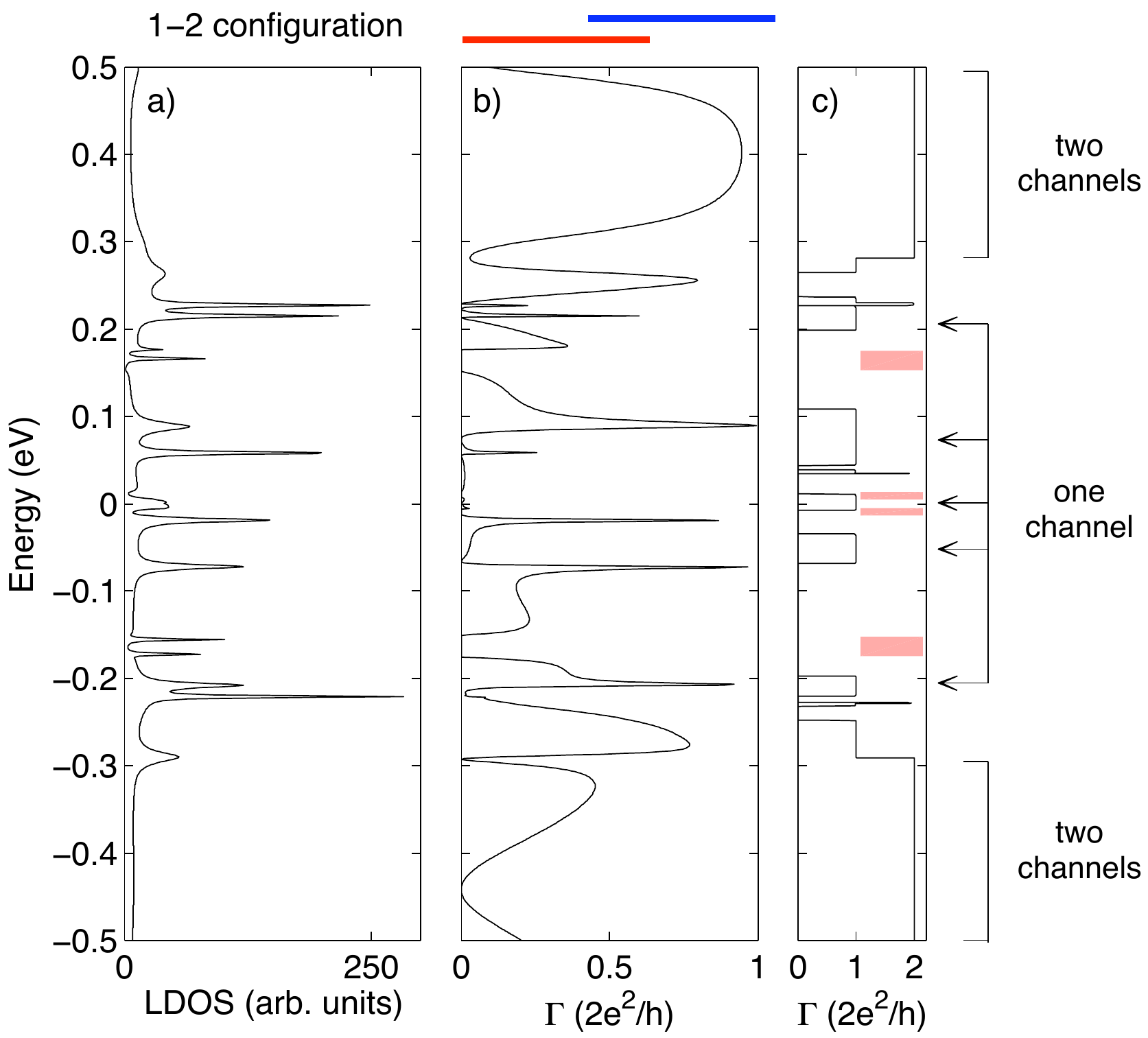}
\caption{\label{fig:ov_ldos_cond} (Color online). 
(a) LDOS and (b) conductance of an overlap (10,1) [(4,3)] flake (configuration 1-2) with length equal to 2 unit cells. (c) Conductance 
of the corresponding infinite twisted bilayer (10,1) [(4,3)] ribbon, shown for comparison. The pink rectangles mark the conductance gaps of a monolayer ribbon with $\vec{T}=(10,1)$ and width equal to $\vec{t}_1$. Arrows indicate some of the main one-channel energy ranges for the bilayer ribbon. The broader two-channel energy intervals are marked with square brackets.}
   \end{figure}

The results for the 1-2 configuration, two overlapping semiinfinite nanoribbons which yield a bilayer flake, are shown in 
Fig. \ref{fig:ov_ldos_cond}. Although the quantitative values differ, the analysis is similar: LDOS peaks around $E_F$ correspond to quasi-localized states that give rise to conductance maxima, due to the presence of only one available state in the flake. For $|E| \gtrsim 0.3$ eV, there are two available states in the flakes, so quasi-localized states produce an destructive interference giving rise to a conductance antiresonance.

Therefore, in both geometries, for the energy ranges for which the bilayer flake has only one available state, LDOS peaks correspond to conductance maxima. In such cases, the presence of quasi-localized states increases the probability of transmission through the flake. On the contrary, when there are two possible paths through the conductor---also signaled by a LDOS peak which  indicates the presence of quasi-localized states---a zero conductance minimum reveals the interference between those two conduction channels. 

Notice that for low energies ($|E|<0.3$ eV), the conductances of the overlap and island configuration are rather similar, differently to the results for symmetric AA and AB stacked flakes \cite{Gonzalez_2011}. However, for $|E|>0.3$ eV, a certain complementarity is observed, like in the symmetrically stacked flakes. Below we explain this behavior by analyzing the spatial distribution of the states. 

\subsection{\label{sec:st_local}Spatial localization of flake states}

\begin{figure}[thpb]
      \centering
\includegraphics[width=\columnwidth]{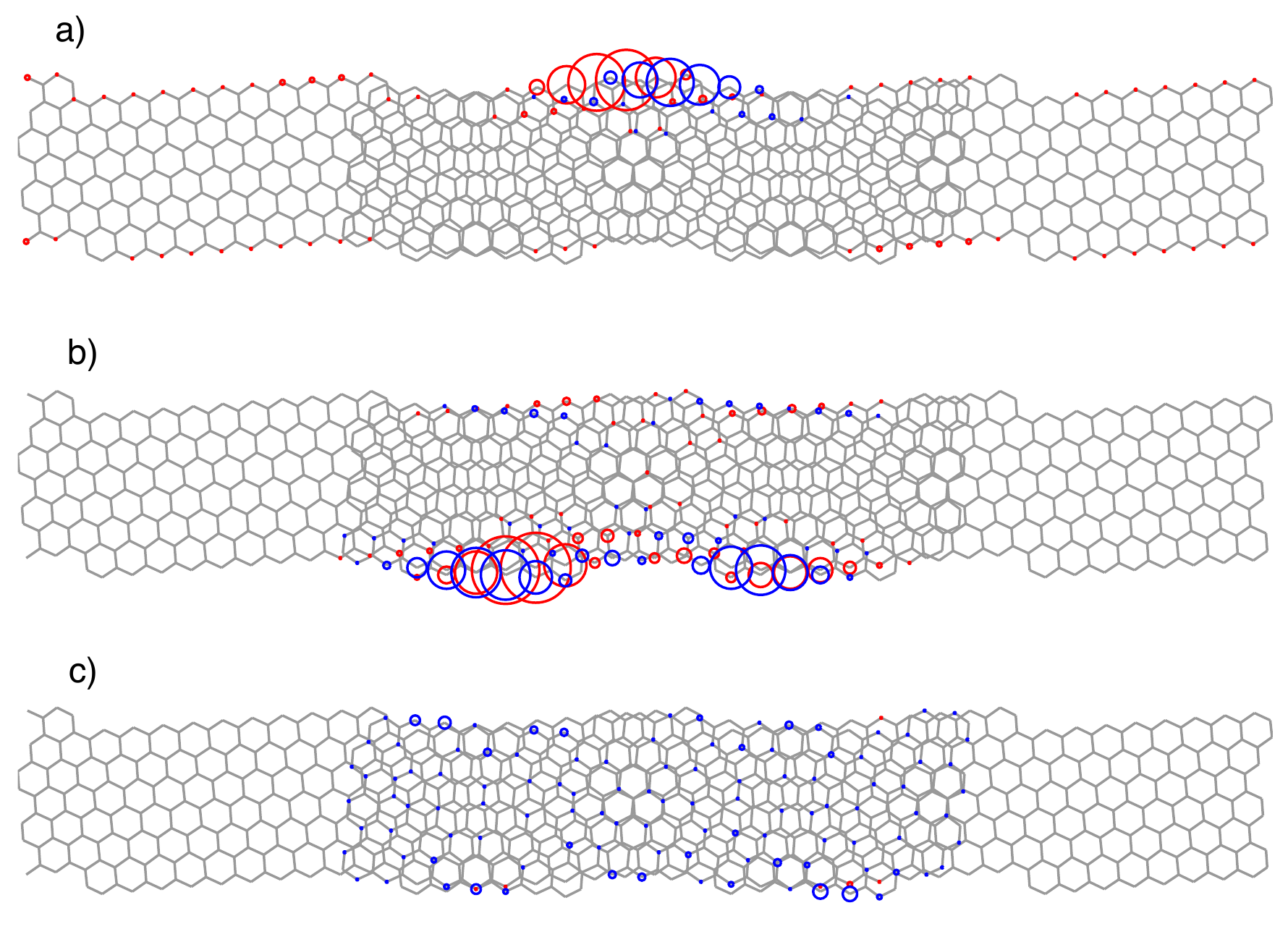}

\caption{\label{fig:island_nodes} (Color online). 
LDOS for three states of 1-1 configuration illustrating different types of localized states. (a) AB-stacking edge state at $E=-0.0186$ eV; (b) AA-stacked edge state at $E=-0.2210$ eV; and (c) flake state at $E=-0.4320$ eV.
 The radii of the circles are proportional to the LDOS value on each atom. Color indicates localization on the bottom (red) and top (blue) layer.}
   \end{figure}

It is interesting to check the spatial localization of the states which give rise to the main conductance features, in order to verify our analysis 
based on the energy dependence of the AA- and AB-stacked edge states. 
As the size of the chosen system is still manageable, we have examined the atom-resolved LDOS for all the quasi-localized states in the energy range $[-0.5,0.5]$ eV. 
Three different examples of localization for the case of the island flake (1-1 configuration) can be seen in 
Fig. \ref{fig:island_nodes}. On the one hand, panel (a) presents a clear example of an AB-edge localized state, with energy $E=-0.0186$ eV. On the other hand, panel (b) is an instance of an AA-localized flake state, with $E=-0.2210$ eV.
The LDOS of these edge-localized states is maximum far from the leads-flake connections. This is the reason why 
the corresponding conductances do not depend much on the configuration of the leads. 
For energies $|E|> 0.3$ eV we observe states related to the finite size of the flake, and indeed they are spread all over the flake region,
mostly in the upper layer 
(Fig. \ref{fig:island_nodes} (c)).
For these cases the conductance does depend on the way the flake is connected to the leads, because the weight of the conductive states is not negligible at the 
junctions between leads and flake.

The differences between the spatial distribution of the states can be related to those in the infinite twisted bilayer nanoribbon. 
For both configurations, states with energies close to the Fermi level ($|E|< 0.15$ eV) are mostly localized on the AB-stacked part of the edges, whereas states that are slightly farther from $E_F$ ($|E|$ around 0.2 eV), have an LDOS concentrated on the AA-stacked edges. 
These edge-localized states are thus far from the leads-flake contacts, so they do not depend much on the leads' configuration. 

\begin{figure}[thpb]
      \centering
\includegraphics[width=\columnwidth]{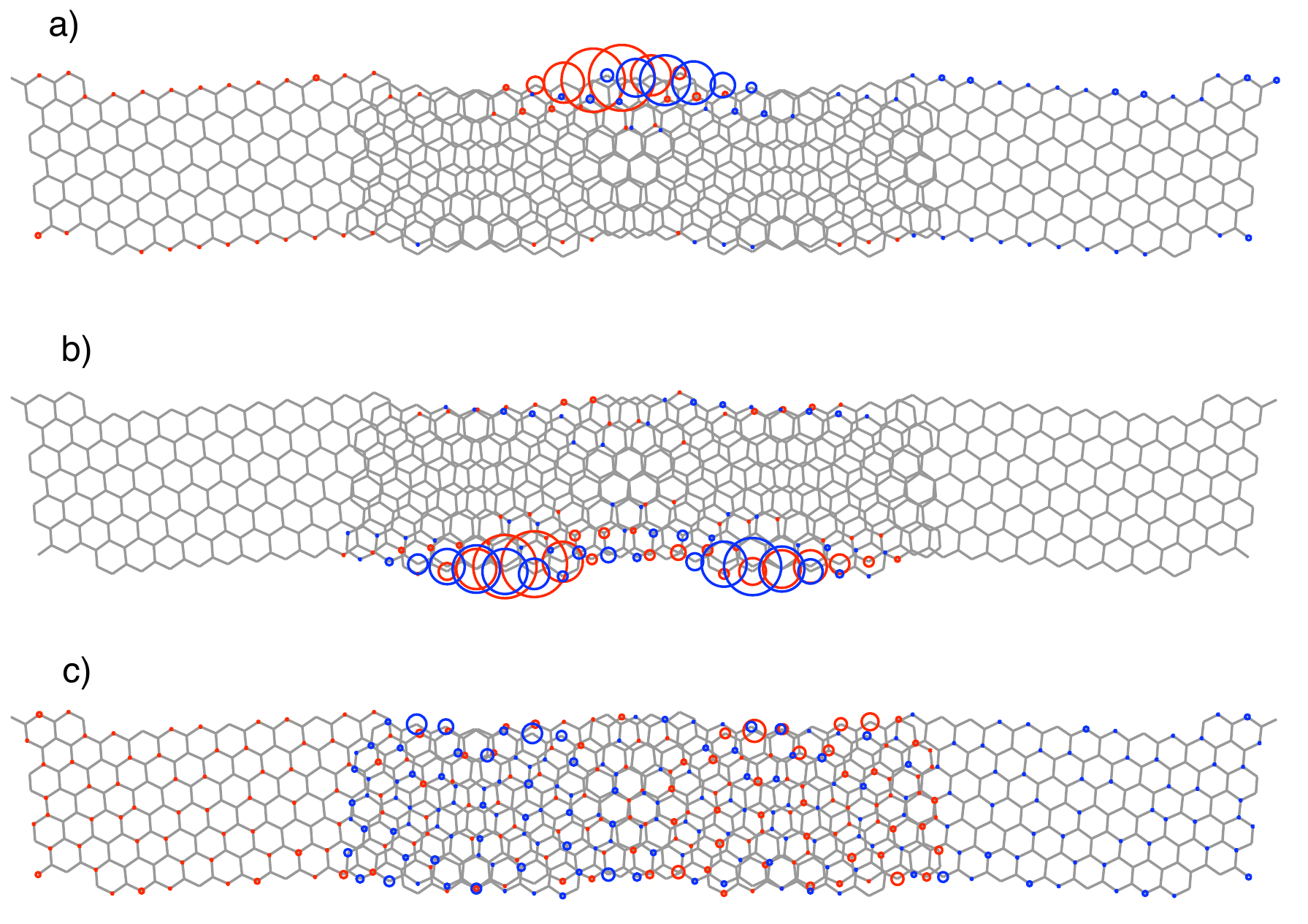}
\caption{\label{fig:overlap_nodes} (Color online). 
LDOS for three states of an overlap flake with different
 localizations. 
 (a)  $E=-0.0195$ eV, edge state with AB stacking; 
(b) $E=-0.2213$ eV, edge state with AA stacking; 
and (c)  $E=0.5025$ eV, flake state.
The radii of the circles
are proportional to the LDOS value on each 
atom. 
In panel (c) LDOS values are multiplied by 20. 
Color indicates 
 localization on the bottom (red) and top (blue) layer.
}
   \end{figure}

When $|E|<0.3$ eV,  a similar behavior is observed for the overlap flake. 
Fig. \ref{fig:overlap_nodes} (a) is an AB-edge state, rather similar to the corresponding case in the island flake, and also with an energy very close to $E_F$ ($E=-0.0195$ eV); it belongs to the AB bands of the infinite bilayer nanoribbon. The same happens to the AA-localized state with energy $E=-0.2213$ eV, 
Fig. \ref{fig:overlap_nodes} (b): in these two cases, confinement is due to the interplay of the moir\'e pattern and the edges, not to the boundaries of the flake, where in fact the LDOS is almost negligible. This explains the similarities in these edge-localized flake states for the two configurations. 
However, the different boundary conditions for the bilayer flake result in distinct behaviors in the overlap and the island flake for energies $|E| > 0.3$ eV 
(Fig. \ref{fig:is_ldos_cond} (b) and 
Fig. \ref{fig:ov_ldos_cond} (b)).
LDOS maxima of these states correspond mainly to flake states, i.e., from quantum size effects for the whole system, not particularly localized at the edges. 
They have a non-negligible weight at the flake boundaries, so they depend on the leads' configuration.
Fig. \ref{fig:overlap_nodes} (c) represents the LDOS for the $E=0.5025$ eV, which corresponds to an antiresonance in the conductance. Here the quasi-localized state is spread all over the flake region.
For energies $|E| > 0.3$ eV, the peaks correspond to Fabry-Perot states in the overlap system 
(Fig. \ref{fig:overlap_nodes} (c)), or quantum dot in the island case, 
(Fig. \ref{fig:island_nodes} (c)), for which localization in the upper flake is stronger. Contrary to the AB and AA-stacked edge localized states, these are only due to the boundary conditions of the flake, i.e., size effects. 
This different type of confinement,
i.e., edge-moir\'e vs. flake size effects, 
 explains the distinct behavior of the conductance for energies closer to $E_F$, related to edges, compared to energies corresponding to flake states which extend over all the system up to the leads, observed in 
 Figs. \ref{fig:is_ldos_cond} and \ref{fig:ov_ldos_cond}. 

Although AB-stacked edge states always lie close to $E_F$, and AA-edge localized states are also roughly split an energy related to the interlayer coupling $|\gamma_1|$,  the rest of energy ranges mentioned above depend on the width of the flake and also of the RRA. For wider ribbons the dispersion of edge states is smaller, and the bulk bilayer nanoribbon bands shift closer to the Fermi energy, so the related flake antiresonances will also occur at smaller energies. 
We have also checked that for wider ribbons localization is still governed by AB-edge states. Eventually, one expects that AA localized regions will 
develop in the inner area of the flake, but for widths around 60 \AA, we still find AB edge-localized states for low energies. 

Larger flakes, i.e., smaller RRA, imply more edge states for the chiralities chosen in this work, so the analysis would be lengthier, but their behavior will be the same as reported for the (10,1) [(4,3)] case, with the obvious differences in the specific values of the energies.  

Notice that, even though the antiresonances reported here correspond to flake states, for longer ribbons edge-localized states may also give rise to conductance minima. As discussed before, there are a few narrow two-channel energy windows corresponding to edge bands in which antiresonances may appear for sufficiently large flakes. In spite of these quantitative differences, the overall behavior described here can be easily extrapolated to other geometries.

\subsection{\label{sec:length}Length dependence}
\begin{figure}[thpb]
      \centering
\includegraphics[width=\columnwidth]{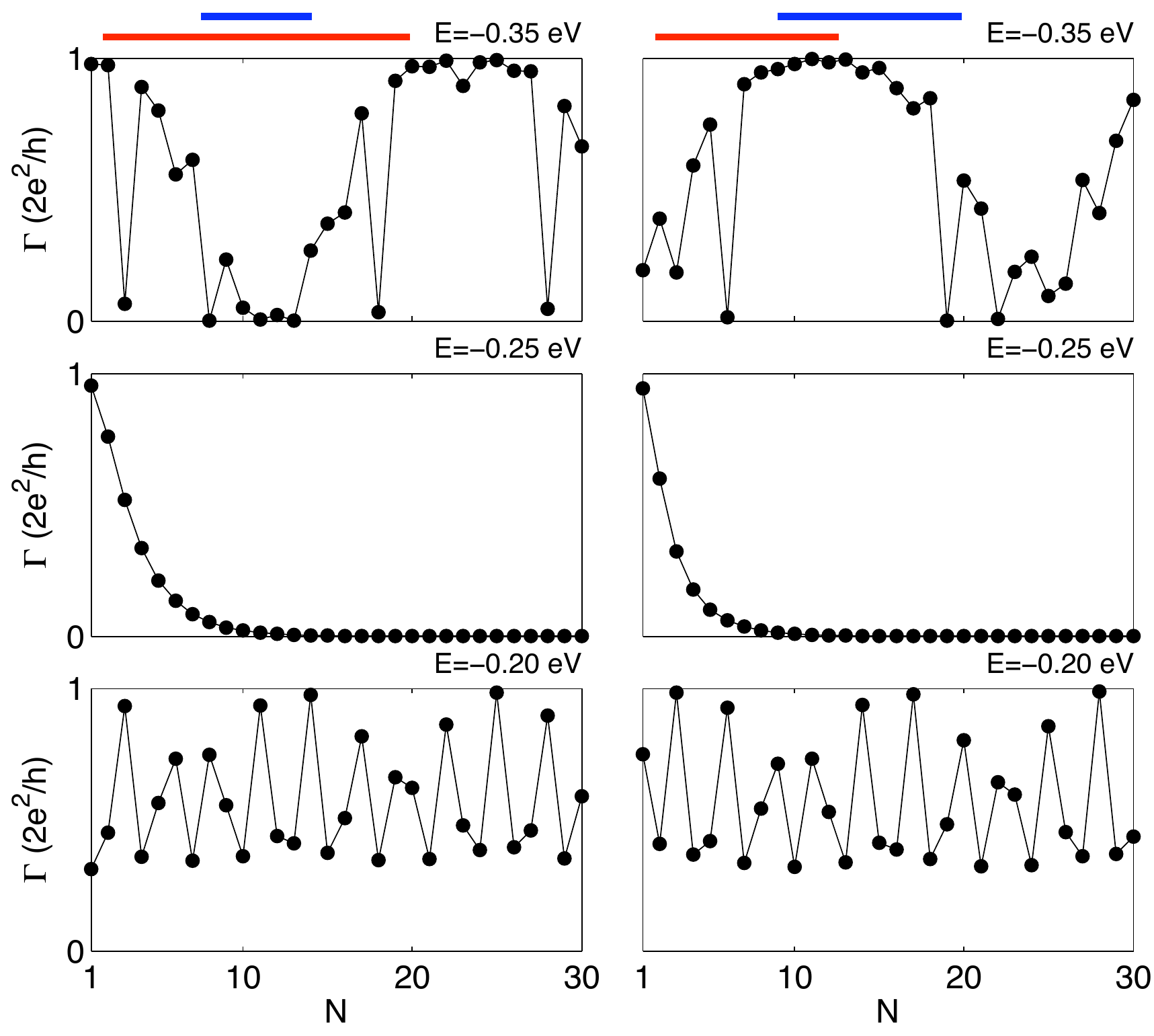}
\caption{\label{fig:osc} (Color online). 
Conductance values as a function of overlap region length ($N$) calculated for both system configurations for three different energy values.}
   \end{figure}

We have also studied the dependence of conductance on the flake length. The results for both configurations are shown on 
Fig. \ref{fig:osc}. 
We select three energy values with distinct transport properties. 
 In the first case,  $E=-0.35$ eV, the bilayer flake has two propagating channels. 

Large-period oscillations can be observed with abrupt antiresonances, characteristic of the two-channel interference. 
  Not all the conductance drops reach zero value due to the discrete sampling of the flake length, $N$ being an integer.
  For the second case, $E=-0.25$ eV, we observe an exponential decay of the conductance. This corresponds to a gap in the bilayer flake spectrum:
for small $N$, there is a non-zero conductance due to tunneling, but for larger flakes transmission cannot occur through the bilayer, which acts as a barrier.
In the third case, $E=-0.20$ eV,  there is a clear oscillatory behavior. However, there are not antiresonances due to the existence of only one transmission channel in the flake, 
so the conductance does not drop below $C_{min}=0.3 \frac{2e^2}{h}$. Obviously, this $C_{min}$ is different for other energy values, like in the symmetric flakes \cite{Gonzalez_2011}. 
 Notice that for $E=-0.35$ eV the conductance in island and overlap configurations has a complementary behavior. Such complementarity was also observed in armchair flakes with AA and AB stackings, most noticeably in the region with two propagating channels \cite{Gonzalez_2011}. In chiral moir\'e flakes, 
 for energies $|E| > 0.3$ eV there are two interfering channels with LDOS spread all over the flake, so the conductance depends on the position of the leads, as in the case of the AA and AB-stacked flakes. The other two cases presented in 
 Fig. \ref{fig:osc} are examples of tunneling through the flake ($E=-0.25$ eV), with exponentially decaying LDOS through the central bilayer region, and of an AA-stacked edge state, ($E=-0.20$ eV), for which the maximum weight of the LDOS is away from the leads-flake connections. 
Therefore, these two latter cases do not depend strongly on the particular lead configuration, so the conductances of the overlap and island geometries are similar. \\

\section{\label{sec:summary} Summary}

We have studied the transport properties of twisted bilayer flakes connected to monolayer leads in two geometries, the island configuration, i.e.,  a monolayer ribbon with a finite-size flake on top, and the overlap configuration, composed of two overlapping monolayer ribbons which yield a twisted bilayer flake. These bilayer flakes are built forming a moir\'e pattern, so different stackings take place at the edges.
 Due to the size of the systems studied, we have employed a tight-binding method with parameters fitted to DFT results for twisted bilayer graphene.  

 The conductance through the flake show resonant transport, tunneling and antiresonances depending on the number of available channels at the bilayer flake. We explain these different features by analyzing the energy spectra of the leads and the bilayer nanoribbon composing the system. 
 
 We have identified two types of localized states: on one hand, due to the moir\'e pattern at the edges, localized edge states with maximum LDOS in the AA and AB regions appear. Because of their spatial localization, far from the flake boundaries and nanoribbon contacts, these edge-localized states present a conductance which does not depend appreciably on the configuration of the leads. This behavior corresponds to energies $|E| < 0.3$ eV. In general, the low-energy transport properties are governed by the AB-stacked edge states. Further away from the Fermi level, we observe states localized on the AA-stacked regions of the edges. On the other hand, at higher energies, $|E| > 0.3$ eV, localized states arise from quantum size effects related to the flake boundaries, so they spread over the whole flake into the nanoribbon contacts. For this reason, such high energy states show a strong dependence on the configuration of the leads, more similar to the symmetrically stacked bilayer ribbons.

 Therefore, for the low-energy transport regime, twisted bilayer flakes show a unique behavior, qualitatively different to that of the symmetric stacking cases: 
their transport properties are determined by the interplay between the moir\'e pattern and the graphene edges.
 Further work considering the role of dispersive forces, e.g., by means of a van der Waals functional \cite{Dion2004, Grimme2006} in twisted ribbons will certainly be of great interest. 

\begin{acknowledgement}

This work was partially supported by the Polish Ministry of Science and Higher Education (The "Mobility Plus" Program) and the Spanish Ministry of Economy and Competitiveness under Grant No. FIS2012-33521. E.S.M. acknowledges FONDECYT grant 11130129.

\end{acknowledgement}

\bibliography{tblgnr_qd_bib}

\end{document}